\documentclass[sigconf]{acmart}
\copyrightyear{2024}
\acmYear{2024}
\setcopyright{rightsretained}
\acmConference[ICSE '24]{2024 IEEE/ACM 46th International Conference on Software Engineering}{April 14--20, 2024}{Lisbon, Portugal}
\acmBooktitle{2024 IEEE/ACM 46th International Conference on Software Engineering (ICSE '24), April 14--20, 2024, Lisbon, Portugal}\acmDOI{10.1145/3597503.3623347}
\acmISBN{979-8-4007-0217-4/24/04}

\newboolean{showcomments}
\setboolean{showcomments}{false}

\usepackage[utf8]{inputenc}
\usepackage{url}
\usepackage{colortbl}
\usepackage{balance}
\usepackage{xspace}
\usepackage{graphicx}
\usepackage{verbatim}
\usepackage{bold-extra}
\usepackage{paralist}
\usepackage{multirow}
\usepackage{listings}
\usepackage{boxedminipage}
\usepackage{soul}
\usepackage{enumitem}
\usepackage[normalem]{ulem}
\setlist{nolistsep,leftmargin=.5cm}
\usepackage{booktabs}
\usepackage{tabularx}
\usepackage{svg}
\usepackage{calc}
\usepackage{float}
\usepackage{multirow}
\usepackage[normalem]{ulem}
\useunder{\uline}{\ul}{}
\usepackage[most]{tcolorbox}
\usepackage{caption}
\usepackage{microtype} 
\usepackage{algorithmic}
\usepackage{graphicx}
\usepackage{textcomp}
\usepackage{xcolor}
\usepackage{hyperref}
\usepackage{ifthen}
\usepackage{wrapfig}
\usepackage{subcaption}
\usepackage{cleveref} 

\ifthenelse{\boolean{showcomments}}
{\newcommand{\nb}[2]{
		\fbox{\bfseries\sffamily\scriptsize#1}
		{\sf\small$\blacktriangleright$\textit{#2}$\blacktriangleleft$}
	}
	
}
{\newcommand{\nb}[2]{}
	
}

\newcommand\os[1]{{\color{red} \nb{OSCAR}{#1}}}

\newcommand\denys[1]{{\color{blue} \nb{DENYS}{#1}}}

\newcommand\rev[1]{{\color{black}{#1}}}
\newcommand\finalrev[1]{{\color{black}{#1}}}

\newcommand{\ie}{\textit{i.e.},\xspace}
\newcommand{\eg}{\textit{e.g.},\xspace}
\newcommand{\etc}{\textit{etc.}\xspace}
\newcommand{\etal}{\textit{et al.}\xspace}
\newcommand{\aka}{\textit{a.k.a.}\xspace}

\begin{abstract}
Software Bills of Materials (SBOMs) have emerged as tools to facilitate the management of software dependencies, vulnerabilities, licenses, and the supply chain. While significant effort has been devoted to increasing SBOM awareness and developing SBOM formats and tools, recent studies have shown that SBOMs are still an early technology not yet adequately adopted in practice.
Expanding on previous research, this paper reports a comprehensive study that investigates
the current challenges stakeholders encounter when creating and using SBOMs.
The study surveyed 138 practitioners belonging to five stakeholder groups (practitioners familiar with SBOMs, members of critical open source projects, AI/ML,  cyber-physical systems, and legal practitioners) using  differentiated questionnaires, and interviewed 8 survey respondents to gather further insights about their experience. We identified 12 major challenges facing the creation and use of SBOMs, including those related to the SBOM content, deficiencies in SBOM tools, SBOM maintenance and verification, and domain-specific challenges. 
We propose and discuss 4 actionable solutions to the identified challenges and present the major avenues for future research and development. 
\end{abstract}

\title{BOMs Away! Inside the Minds of Stakeholders:\\A Comprehensive Study of Bills of Materials for Software Systems}

\author{Trevor Stalnaker}
\email{twstalnaker@wm.edu}
\affiliation{
  \institution{William \& Mary}
  \city{Williamsburg}
  \state{Virginia}
  \country{USA}
}

\author{Nathan Wintersgill}
\email{njwintersgill@wm.edu}
\affiliation{
  \institution{William \& Mary}
  \city{Williamsburg}
  \state{Virginia}
  \country{USA}
}

\author{Oscar Chaparro}
\email{oscarch@wm.edu}
\affiliation{
  \institution{William \& Mary}
  \city{Williamsburg}
  \state{Virginia}
  \country{USA}
}

\author{Massimiliano Di Penta}
\email{dipenta@unisannio.it}
\affiliation{
  \institution{University of Sannio}
  \city{Benevento}
  \country{Italy}
}

\author{Daniel M German}
\email{dmg@uvic.ca}
\affiliation{
  \institution{University of Victoria}
  \city{BC}
  \country{Canada}
}

\author{Denys Poshyvanyk}
\email{denys@cs.wm.edu}
\affiliation{
  \institution{William \& Mary}
  \city{Williamsburg}
  \state{Virginia}
  \country{USA}
}

\begin{document}



\maketitle

\sloppy

\section{Introduction}

The software supply chain has increasingly grown in complexity with the proliferation of open-source software \cite{growing_complexity,challengesSecuringSupplyChain} and AI/ML components \cite{tan2022exploratory, jiang2023empirical, jiang2022empirical}. Organizations and developers \rev{often accomplish their tasks by integrating components from a variety of vendors \cite{sbom_important}.}
However, leveraging external packages does not come without a cost. The fate of a software product is intrinsically tied to its evolving dependencies \cite{PentaGGA10}. If a dependency 
displays a vulnerability, then so too could the final product, potentially leading to severe  consequences \cite{ohm2020backstabber}. \rev{Moreover, failing to comply with the license terms of software dependencies could lead to severe legal and economic consequences for organizations~\cite{vendome2015license, vendome2017license, gangadharan2012managing, riehle2019open}.} 

In this scenario, Software Bills of Materials (SBOMs) have emerged as mechanisms that facilitate the management of software dependencies \cite{ntiaGlance}, leading to improved management of software vulnerabilities, enhanced license compliance, and increased transparency in the software supply chain \cite{ntia_benefits}. 

While SBOMs were introduced in the early 2010s \cite{spdx_history}, the 2021 US Presidential Executive Order 14028 on Improving the Nation's Cybersecurity \cite{execOrder} gave new momentum to SBOM formalization and adoption \cite{supplyChainRiskManage} as it required companies selling software to the US government to provide SBOMs. This was prompted by recent supply chain attacks, such as the SolarWinds breach~\cite{peisert2021perspectives} and critical vulnerabilities such as those affecting the Log4J library~\cite{log4j}, which impacted many users \cite{log4j_CVE,log4j_google}. SBOMs are currently championed by the US National Telecommunications and Information Administration (NTIA) 
\cite{ntiaGlance,ntiaFraming} and  well-known organizations such as the Linux Foundation \cite{linux_foundation} and OWASP \cite{owasp}. Significant effort has been put into  promoting SBOM formats and tools that can create and process SBOMs \cite{ntia_tooling}, with the goal of increasing adoption and fully enabling the benefits that SBOMs offer~\cite{ntia_benefits}.



Although organizations and developers have acknowledged the importance of SBOMs and anticipate using them more frequently in the coming years \cite{sboms_matter,linuxSbomSurvey}, recent research highlighted concerns regarding their commitment to SBOMs and the actual benefits SBOMs bring to their projects \cite{xia2023empirical,linuxSbomSurvey,linuxSbomSurvey}. These concerns arise due to the lack of industry agreement regarding the content of SBOMs across different domains, as well as how they should be employed and integrated into their development and operational processes \cite{linuxSbomSurvey,iti}. 
An additional barrier is the lack of mature tools for SBOM production and consumption \cite{xia2023empirical,linuxSbomSurvey,zahan2023software}. 


In light of these findings, it is imperative to understand (i) how developers and other stakeholders currently create and use SBOMs, (ii) additional opportunities/benefits that SBOMs can offer for different types of software and stakeholders, (iii) the specific challenges that prevent stakeholders from fully exploiting the SBOM benefits, and (iv) actionable solutions to overcome such challenges.


This paper contributes to the body of knowledge about SBOM adoption by reporting a comprehensive empirical investigation of the aforementioned aspects. The study combined survey questionnaires with semi-structured interviews. Given the diverse types of modern software systems which SBOMs should support, 
we distributed five distinct questionnaires to different groups of stakeholders, 
resulting in a total of \rev{150} responses (\rev{84} responses indicating SBOM familiarity). Specifically, the surveys targeted software practitioners familiar with SBOMs, contributors of critical open-source systems (OSS) \cite{100_critical_projects}, AI/ML, Cyber-Physical Systems (CPS) \cite{carmody2021building}, and legal practitioners. To gain a deeper understanding of the key SBOM experiences, opportunities, challenges, and solutions collected in the surveys, we conducted semi-structured interviews with eight participants from different groups. 
\looseness=-1



Our study expands our understanding of the  limitations and challenges of SBOM formats and tools,  identifies areas that research and practice on SBOM support should focus on, and provides a thorough discussion of potential solutions to overcome these barriers.
\looseness=-1




In summary, the main contributions of this paper are:
\begin{itemize}
    \item An empirical study of SBOM adoption, challenges, and solutions. The study targeted five groups of stakeholders, according to  the different types of software that SBOMs should support, offering greater scope and different perspectives about SBOM adoption compared to recent prior studies \cite{xia2023empirical, balliu2023challenges,zahan2023software}; 
    
    \item A deep analysis and discussion of how software stakeholders use and create SBOMs, new opportunities/benefits that SBOM can offer, and  challenges that prevent stakeholders from fully exploiting the SBOM benefits; and
    \item A thorough discussion and proposal of actionable solutions for the identified challenges and obstacles, as well as key areas that researchers and practitioners should focus on to improve SBOM production and consumption.
\end{itemize}



\section{Background and Related Work}
\label{sec:related}



Bills of Materials (BOMs) 
refer to the list of raw materials, components, and parts needed to manufacture an end product~\cite{harvardreview, rusk1990role}. The concept has been transferred to software systems as Software BOMs (SBOMs)\rev{, which identify a project's dependencies and their provenance.}  Three major SBOM format specifications currently exist: SPDX \cite{spdx_spec}, CycloneDX \cite{cycloneSpecs}, and SWID \cite{swid_specification}. While the NTIA has not officially endorsed any one specification \cite{surveyOfSBOMStandards}, SPDX was officially recognized as a standard by ISO in 2021 \cite{spdx_iso_spec}.
\looseness=-1

\finalrev{Software component inventory, vulnerability analysis, and license compliance are primary SBOM use cases~\cite{cyclonePractical}.  SPDX, supported by the Linux Foundation~\cite{linux_foundation}, began as a solution for managing open-source licenses and later became an SBOM standard for documenting software components, licenses, security-related information, and other metadata. CycloneDX, supported by OWASP~\cite{owasp}, provides virtually the same features as SPDX, but focuses primarily on security and vulnerability management.
\looseness=-1

Both specifications support several file formats. SPDX: tag/value (.spdx), JSON, YAML, RDF/XML, and spreadsheets (.xls) \cite{spdxFormats}; CycloneDX: JSON, XML, and ``protocol buffers''~\cite{cycloneFormats}.  Example SBOMs for each format can be found at ~\cite{cyclone_sbom} and ~\cite{spdx_sbom} respectively.
\looseness=-1

The differing design philosophies result in a few notable differences~\cite{cyclonePractical, cycloneVsSPDX}. Unlike CycloneDX, SPDX can represent code snippets within files (and their licenses), and supports annotations (adding comments to an SPDX document, \eg clarifications about ambiguous legal content). CycloneDX natively supports ``compositions'', which allow expressing the completeness level of a BOM element (\eg dependency relationships)---SPDX does not support this feature directly (only through annotations). CycloneDX offers more robust support for vulnerability management. For example, CycloneDX allows software suppliers to assert software vulnerabilities via the Vulnerability Exploitability eXchange (VEX) format, which SPDX does not support.

}

As modern software systems go beyond the mere integration of libraries and frameworks, various initiatives have proposed different types of BOMs, to account for other components typically integrated into a software system (\eg hardware devices, firmware,  APIs, or AI/ML models). Practitioners have proposed BOMs for:

\begin{itemize}[topsep=0pt,parsep=0pt,partopsep=0pt]
    \item 
external services/APIs (SaaSBOMs)~\cite{saas_governance, cyclone_saas_bom, reasons_for_saasbom}; 
    \item 
hardware (HBOMs)~\cite{cyclone_hbom}  and firmware (FBOMs)~\cite{fbom_presentation,secure_iot_fbom,firmware_security_fbom};
    \item operational (\eg configuration) environments (OBOMs)~\cite{obom}; \&
    \item datasets (DataBOMs)~\cite{barclay2019towards} and AI models (AIBOMs)~\cite{first_aibom_mention,xia2023empirical}. 
\end{itemize}
\looseness=-1

Our study targets specific populations of software stakeholders (\eg AI and Cyber-Physical Systems practitioners) to understand needs that could be potentially fulfilled by various kinds of BOMs.
\looseness=-1



While SBOMs have existed for some time \cite{spdx_history, cyclonedx_history, swid_specification, history_of_sbom}, they are only now beginning to be widely known \cite{considerable_progress, ntia_progressing_well}. 
The analysis of their uses and shortcomings has been investigated only by a few recent studies \cite{martin2020visibility, xia2023empirical, balliu2023challenges, carmody2021building, linuxSbomSurvey, caven2022integrating, tang2022towards,zahan2023software}, which we discuss next.
\looseness=-1



A survey from the Linux Foundation examined the current state of SBOM usage and readiness in industry \cite{linuxSbomSurvey}, aiming to identify the main use cases, benefits, and unmet needs for SBOMs. The study examined SBOM adoption, claiming that of 400 organizations surveyed worldwide, an estimated 78\% would use SBOMs by 2022 and 88\% by 2023. 
Our work differs in that we seek to identify the SBOM usage needs of developers, not organizations.
\looseness=-1

Caven \etal surveyed US Department of Defense officials to examine what features they look for when making procurement decisions, including features that are part of SBOMs \cite{caven2022integrating}. They found that, generally, source code used in development was the least-important feature to include in SBOMs, and SBOMs were valued differently by people of different roles. In contrast to their survey, our study investigates the adoption of SBOMs from different perspectives, by targeting different sub-populations of stakeholders via distinct questionnaires and follow-up interviews.
\looseness=-1

In a study on C/C++ libraries \cite{tang2022towards}, Tang \etal found little evidence of SBOMs being used in open-source projects. Of 24K+ GitHub repositories examined, fewer than ten contained recognizable SBOMs, yet some use package manager files, which provide similar information as SBOMs (\ie they are "quasi-SBOMs").

\rev{The gray literature review by Zahan \etal examined common benefits and challenges of adopting SBOMs~\cite{zahan2023software}. Benefits include enhanced dependency, vulnerability, risk, and licensing management, and better competitive advantage. Challenges include the lack of SBOM tooling, interoperability, and value, as well as extra effort and disclosure of sensitive information.  
In our work, we study how these benefits and challenges are perceived by different stakeholder groups, who use a variety of  software and BOM types.}
\looseness=-1


Xia \etal interviewed 17 software practitioners to derive 25 statements about SBOM practices, tools, \& concerns. They surveyed 65 practitioners, who indicated their agreement with the statements 
and commented on  
their experience with SBOMs.
Ten findings were derived from their responses, including the need to integrate SBOM formats to support various usage scenarios (\eg including vulnerability data), limited level of SBOM awareness, immaturity of SBOM tools, and lack of suitable trust mechanisms. Our study extends this prior work as it: (1) includes five surveys that target diverse stakeholders (\eg AI/ML, CPS, and legal practitioners), (2) investigates usage, challenges, \& solutions for different BOMs and software types, (3) analyzes the specific challenges of creating and using SBOMs, and (4) discusses solutions to overcome these challenges.
\looseness=-1



Lin \etal explored the use of SBOM tools for DevSecOps and software composition analysis \cite{lin2023generating}. 
Balliu \etal compared six state-of-the-art tools that generate SBOMs for Java systems and compared how accurate the SBOMs are in listing project dependencies, compared to those given by Maven~\cite{balliu2023challenges}. The tools capture a different set of project dependencies, missing much of the Maven dependency tree. 
While Balliu \etal discuss open challenges for accurate/effective SBOM generation  and usage,
our study provides a more comprehensive view of different stakeholders' problems regarding SBOMs, beyond Java systems, SBOM tools, and security-related applications. 
\looseness=-1



There are different proposals to track datasets and AI model information \cite{bender2018data, holland2018dataset, mitchell2019model, gebru2021datasheets}.  None of them apply the concept of BOMs to data/model supply chains.  The term DataBOM was introduced and discussed by Barclay \etal~\cite{barclay2019towards} without, however, surveying developers to investigate its feasibility.  Potential use cases for the AI/ML domain are mentioned, but DataBOMs are never considered within the context of AIBOMs. In our study, we ask stakeholders about the potential relationship between DataBOMs and AIBOMs.
\looseness=-1

The concept of AIBOM was proposed by Chan in 2017~\cite{first_aibom_mention}, but no specific implementation details or recommendations were given.  Barclay \etal, building on their previous work, explored how SBOMs might be applied in the context of AI/ML systems \cite{barclay2022providing}.

\begin{table}[t]
\centering
\caption{Methodology and scope of SBOM studies}
\resizebox{\columnwidth}{!}{%
\label{tab:methodology_comparison}
\begin{tabular}{llll}
\textbf{Study} & \textbf{Research methods} & \textbf{Considered BOMs} & \textbf{Study participants} \\ \hline
Boms Away &

\begin{tabular}[c]{@{}l@{}}
Five surveys and \\
follow-up interviews 
\end{tabular} 
&

\begin{tabular}[c]{@{}l@{}}
SBOMs, HBOMs, \\
AIBOMs, \& DataBOMs
\end{tabular} 
 &

\begin{tabular}[c]{@{}l@{}}
SBOM Producers, \\ Consumers,
Tool Makers, \\ Standard Makers,
and Educators;\\ 
Developers of Critical OSS projects;\\ 
AI/ML, CPS, \& Legal \\ practitioners/researchers
\end{tabular} 

\\ \hline
Xia \etal's~\cite{xia2023empirical} &

\begin{tabular}[c]{@{}l@{}}
Interviews to derive \\
one survey 
\end{tabular} 

& SBOMs \& AIBOMs & Developers \\ \hline
Zahan~\etal's~\cite{zahan2021weak} & Grey literature review & SBOMs & -
\\ \hline
Linux Found.~\cite{linuxSbomSurvey} & One survey & SBOMs & Software organizations
\end{tabular}%
}
\vspace{-.2cm}
\end{table}
\finalrev{\Cref{tab:methodology_comparison} provides a comparison between our study and the most related prior studies, regarding methodology \& scope.  A more detailed comparison can be found in our replication package~\cite{anonymous_repo}.}




\begin{table*}[t]
\caption{Survey questions for different participant groups}
\label{tab:survey}
\setlength{\tabcolsep}{2pt}
\resizebox{\textwidth}{!}{
\begin{tabular}{ll}
\hline
\textbf{Survey Group}                         & \textbf{Question Topics}                                                                                                             \\ \hline
SBOM Community and Adopters                   & SBOM content and use cases, SBOM benefits/challenges, SBOM usage for security (by role: consumers, producers, \etc)  \\
Contributors of Critical OSS  & SBOM content and use cases, OBOM content and adoption, other BOM practices                                                           \\
AI/ML Developers/Researchers       & AIBOM content and use cases, DataBOM content and use cases, benefits/challenges of BOMs for AI/ML                            \\
CPS Developers/Researchers & SBOM content and use cases, HBOM content and adoption, benefits/challenges of BOMs for CPSs  \\
Legal Practitioners                  & SBOM requirements, SBOMs in legal agreements, software licensing, DataBOM use cases                               \\ \hline
\end{tabular}
}
\vspace{-0.28cm}
\end{table*}

\section{Study Design}

The \emph{goal} of our study is to investigate the challenges encountered by stakeholders when creating and using SBOMs, and how such challenges can be addressed. The \emph{context} of the study consists of five stakeholder groups: software developers, project leaders and contributors, AI/ML, CPS, and legal practitioners.

\begin{figure}[t]
\centering
\includegraphics[width=1\linewidth]{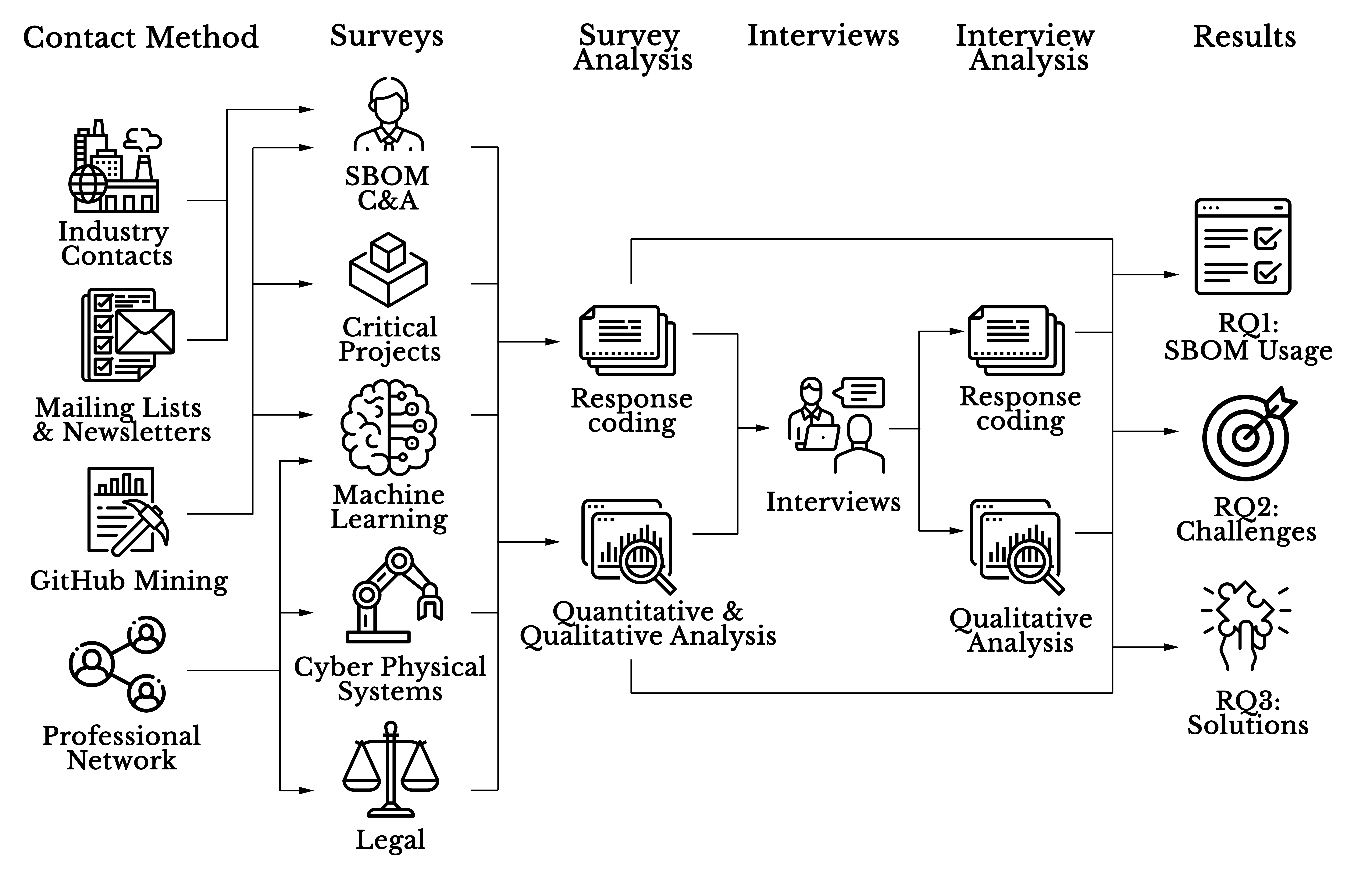}
\caption{Research methodology 
\small (image credits at \cite{anonymous_repo})}
\label{fig:methodology}
\vspace{=-0.3cm}
\end{figure}


The study aims to address the following research questions (RQs):

\begin{enumerate}[label=\textbf{RQ$_\arabic*$:}, ref=\textbf{RQ$_\arabic*$}, wide, labelindent=5pt]\setlength{\itemsep}{0.2em}
    \item \label{rq:usage}{\textit{How do software stakeholders create and use SBOMs?}}
    \item \label{rq:challenges}{\textit{What are the challenges of creating and using SBOMs?}}
    \item \label{rq:solutions}{\textit{What are actionable solutions to SBOM challenges?}}
\end{enumerate}

We next describe the study methodology to answer the RQs, which includes five distinct surveys and follow-up interviews with participants from different stakeholder groups (\cref{fig:methodology}).

As the study involves human subjects, the methodology (including procedures to gather contact information, recruitment methods, survey/interview questions and format, data analysis, and dissemination methods) has been approved by the ethical board
of the University directly involved in running the study.

\subsection{Survey Design}

Considering the study goal and the RQs, we have designed the survey questionnaires considering previous literature on SBOMs described in Section \ref{sec:related}, general guidelines for survey design \cite{survey}, as well as SE specific guidelines \cite{DBLP:journals/sigsoft/PfleegerK01,DBLP:journals/sigsoft/KitchenhamP02,DBLP:journals/sigsoft/KitchenhamP02a,DBLP:journals/sigsoft/KitchenhamP02b,DBLP:journals/sigsoft/KitchenhamP02c,DBLP:journals/sigsoft/KitchenhamP03}.

Since the study foresees the involvement of a general population of: (1) software developers and other stakeholders that have interacted with SBOMs, and (2) domain specialists (AI/ML, CPSs, and legal practitioners), we designed questionnaires with  questions asked to all stakeholder groups and questions asked to specific groups. 
\looseness=-1

\Cref{tab:survey} summarizes the information we asked for in the surveys. A detailed description of all questions can be found in our replication package \cite{anonymous_repo}. 
The surveys contain a mix of (five-point) Likert-scale, multiple-option, and open-ended questions that asked about: SBOM content,  use cases, benefits, distribution preferences, challenges, potential solutions, dependency management practices, and legal aspects.
All questionnaires also featured a consent form, a statement about data confidentiality, and a demographics section (asking about professional role, software domains, education, known programming languages, and knowledge about software security and licensing).  \rev{Participants who completed the survey entered into a lottery to win one of ten \$50 USD Amazon gift cards.}


\vspace{1cm}
\subsection{Participant Identification}
\label{sec:context}


\rev{To explore different facets of SBOMs and their usage, we identified five participant groups:  SBOM Community and  Adopters, contributors of critical OSS~\cite{100_critical_projects}, as well as AI/ML, 
Cyber-Physical Systems (CPS), and legal practitioners.}
\looseness=-1


\noindent\textbf{SBOM Community and Adopters (SBOM C\&A).}
These are people who work with SBOMs in different manners \cite{arora2022strengthening,carmody2021building}. Contacting people who directly use SBOMs and related technologies allowed us to obtain firsthand feedback on how SBOMs are currently used, as well as any perceived deficiencies in current SBOM standards and tools.  
Within this group, we identified five sub-groups of stakeholders. While we did not explicitly categorize individual stakeholders when selecting potential participants, we asked the participants to self-identify as belonging to \rev{one or more} of the following groups:
\begin{itemize}[topsep=0pt,parsep=0pt,partopsep=0pt,wide]
\item \textbf{SBOM Consumers:}
People who read an existing SBOM to gather information about dependencies, vulnerabilities, or licenses.
\item \textbf{SBOM Producers:}
People who document a software system and its dependencies in an SBOM using a particular format (\eg SPDX, CycloneDX, or SWID).  
\item \textbf{SBOM Tool Makers:}
People who contribute to the development of tools that facilitate the creation or use of SBOMs, \eg SBOM generators from project build scripts or dependencies. 
\item \textbf{SBOM Educators:}
People who create or compile educational resources about SBOMs, including guides and tutorials.
\item \textbf{SBOM Standard Makers:}
People who contribute to specifications for the creation and usage of SBOMs.  These individuals may come from government agencies, corporations, or academia.
\end{itemize}

Eligible participants for this group have been identified based on their potential experience with SBOMs, the supply chain, and software development, via a combination of three different approaches:
\begin{enumerate}[topsep=0pt,parsep=0pt,partopsep=0pt,wide]
 \item \underline{Keyword-based search of GitHub repositories.}
Combining manual effort and automated tools (based on GitHub APIs \cite{gitAPI}), we located public GitHub repositories by searching issues, commits, and files for keywords and traces related to SBOMs and the supply chain. We identified contributors who may have worked with SBOMs by locating repositories with SBOM-related files (\eg associated with the SPDX, CycloneDX, and SWID formats). From these repositories, we mined relevant commits, matching  keywords such as "SBOM," "SPDX," and “bill of materials”. 
From the matched commits, we gathered only publicly available contact information.
A similar approach to identify participants was used by Xia \etal~\cite{xia2023empirical}. 
\looseness=-1


\item \underline{Identifying dependencies between GitHub repositories.}
We found extra eligible participants by (i) examining GitHub profiles/organizations that listed projects with SBOM-related tags as topics, and (2) using GitHub's dependency feature \cite{dependencyGraph} to locate dependent projects with SBOM-related tags.  These repositories and their contributors logically represent groups currently using SBOMs.  \rev{In total, we identified 4,423 developer email addresses via GitHub mining.}

\item \underline{Sharing the survey in relevant mailing lists.} To locate additional individuals familiar with SBOMs, we published a call for participants through SBOM-related mailing lists, including the SPDX~\cite{spdx_mailing_list} and the OpenChain mailing lists \cite{openchain_mailing_list}. 

\end{enumerate}

\noindent\textbf{Developers of Critical Open Source Systems.}
The Open Source Software Foundation's workgroup on Securing Critical Projects compiled a list of the 102 most critical OSS, comprising 564 total repositories \cite{100_critical_projects}. The projects include the Linux Kernel, programming languages, package managers, build systems, databases, \etc 
~
\rev{Given the role of SBOMs in the software supply chain, we sought to administer a targeted survey examining these critical projects, which are widely depended on and may have a greater need to produce, use, and distribute SBOMs.  The actions of these projects are also likely to represent and set the tone for the rest of the open-source landscape. Also, examining these projects allowed us to assess how SBOMs have spread beyond early adopters.} 

Using the GitHub API, we mined the top-10 contributors (by \# of commits) for each of these 564 repositories.  Where there were fewer than ten total contributors, we examined all that were available. 

\noindent\textbf{CPS Developers and Researchers.}
\rev{These are people with expertise in cyber-physical systems (autonomous vehicles, medical monitoring and
industrial control systems, robots \etc), which entail a close interaction between hardware and software.  Given these systems have their own supply chains and are becoming more popular in certain domains, surveying this group allowed us to examine unique challenges facing the usage of SBOMs and HBOMs, as well as how the two may interact.  CPS participants were identified from our professional network.}

\noindent\textbf{AI/ML Developers and Researchers.} These are: (i) Top-10 (by number of commits) developers that contribute to a machine learning project hosted on GitHub \rev{(with 100+ stars)} and expose a public profile. AI/ML projects were identified by matching the projects' topics to keywords such as "machine learning" or "artificial intelligence" (see the full list of keywords in our replication package \cite{anonymous_repo}); and (ii) AI/ML practitioners in our academic/professional network. 

\rev{AI/ML components have their own supply chains, but are also increasingly integrated into traditional software products. Model/data provenance is essential to security (\eg model poisoning), licensing, usage, and research of AI/ML systems.  The needs, challenges, and use cases facing AI/ML developers may be similar and different from those of typical SBOM users. By surveying this group, we aimed to understand these similarities and differences.}

\noindent\textbf{Legal Practitioners.} 
Through our professional network, we identified a legal practitioner with a technical background who could answer questions about non-technical challenges facing SBOM use. This includes examining how SBOMs interact with regulations, contractual obligations, and more. \rev{The views of one respondent are not representative of the field at large, but with only a small pool of legal practitioners having software development and SBOM  experience, this group is the hardest group to survey at scale.}


\vspace{-0.28cm}
\subsection{Survey Response Collection and Analysis}

Survey responses were collected using Qualtrics~\cite{qualtrics}. Survey participants were only presented with questions related to the group(s) they selected. The survey for SBOM community and adopters was kept open for four months, with three waves of invitations. The remaining surveys were kept open for two to four weeks.

Via email and mailing list posts, \rev{we invited 4.4k+} individuals to participate in the surveys and received \rev{229} complete responses in total (see \Cref{tab:response_stats}). After removing personal information, the responses were analyzed following the procedure described below, resulting in \rev{150} valid responses. \Cref{tab:participants} overviews the demographics for all the study participants.





For the closed-ended questions, we aggregated results using descriptive statistics and discussed them. In particular, we examined responses from Likert-scale questions to determine practitioner sentiments, as well as frequently-selected answers to multiple-choice questions to identify common SBOM use cases and challenges. We report the most frequently selected answers in \Cref{sec:results}. 

For the open-ended questions, a coding approach was applied in line with \cite{spencer2009card}. Two authors ("annotators" in the following) performed a first phase of \textit{open coding} on the first 28 valid responses of \rev{101} received for the SBOM community and adopters survey. They independently assigned one or more codes to each response.   

Once both annotators completed the open coding for the first 28 valid responses, they convened to settle disagreements and consolidated a set of labels. Since multiple codes could be assigned to each response and disagreements were discussed, we did not base our analysis on inter-rater agreements. 

From this point, the remaining responses were coded by the annotators independently. During the further coding, the annotators started from the previously-established codes (available in a shared spreadsheet); yet, they had the option of adding new codes, that would, in turn, 
become available to the other annotator.

After the coding was completed, annotators met to discuss their coding and reconcile the disagreement cases.  Results were analyzed by leveraging descriptive statistics on the codes the annotators assigned to each question. Our replication package contains the code catalog derived from the analysis for each survey and question, which includes the tag and a brief description of the code~\cite{anonymous_repo}. 

\rev{Throughout the whole coding process, the annotators flagged and reviewed answers that were nonsensical, did not answer the survey questions, were copy-pasted from the web, or appeared to be generated through ChatGPT \cite{chat_gpt}. These were reviewed by (1) inspection and discussion between annotators; (2) searching the response text using Google and validating if the text was found verbatim on the Web; or (3) validating the presence of prose, abnormal wordiness, and unusual markup characteristic of ChatGPT responses. In this way, 41 responses were removed from the analysis. Another 20 responses were removed because of numerous blank or repeated answers, and 18 were discarded as spam (\eg same email/IP addresses or identical responses). 
The annotators examined the survey responses and independently flagged potentially invalid responses. They discussed the cases and reached a consensus on the responses to remove and the main
reason for removal.
}
\looseness=-1


\begin{table}[t]
\caption{Number of Survey Respondents} 
\label{tab:response_stats}
\resizebox{\columnwidth}{!}{
\begin{tabular}{lrrrrllr}
\cline{1-5} \cline{7-8}
\textbf{Survey}   & \begin{tabular}[c]{@{}c@{}}\textbf{Full} \\ \textbf{Resps}\end{tabular} & \begin{tabular}[c]{@{}c@{}}\textbf{Valid} \\ \textbf{Resps}\end{tabular} & \begin{tabular}[c]{@{}c@{}}\textbf{Fam. w/} \\ \textbf{SBOMs}\end{tabular} & \multicolumn{1}{l}{\begin{tabular}[c]{@{}l@{}}\textbf{Inter-}\\ \textbf{views}\end{tabular}} & \multirow{7}{*}{\textbf{}} & \textbf{Role} & \textbf{\#} \\ \cline{1-5} \cline{7-8} 
SBOM C\&A     & 179                                                    & \rev{101}                                                     & 61                                                      & 4                                                                           &                            & P    & \rev{34} \\ 
Critical & 22                                                     & 22                                                     & 13                                                      & 1                                                                           &                            & C    & 31 \\ 
ML       & 21                                                     & 20                                                     & 8                                                       & 1                                                                           &                            & TM   & 24 \\ 
CPS      & 6                                                      & 6                                                      & 1                                                       & 1                                                                           &                            & E    & 14 \\ 
Legal    & 1                                                      & 1                                                      & 1                                                       & 1                                                                           &                            & SM   & 16 \\ 
\cline{1-5} 
\textbf{Total}   & \textbf{229}                                                    & \rev{\textbf{150}}                                                   & \textbf{84}                                                      & \textbf{8}                                                                           &                            & O    & 7  \\ \cline{1-5} \cline{7-8} 
\end{tabular}
}
{\footnotesize P=Producer, C=Consumer, TM=Tool Maker, E=Educator, SM=Std. Maker, O=Other}
\vspace{-0.2cm}
\end{table}

\begin{table}[]
\caption{Abbreviated participant demographics}
\label{tab:participants}
\setlength{\tabcolsep}{2pt}
\resizebox{\columnwidth}{!}{
\begin{tabular}{llll}
\hline
\textbf{Survey}                    & \textbf{Top Software Domains}      & \textbf{Top Roles}               & \textbf{Experience (yrs)} \\ \hline
\multirow{3}{*}{SBOM C\&A}     & Web apps 76\% (38)        & Programmer 30\% (18)    & 0-5 16\% (8)     \\
                          & Desktop apps 56\% (28)    & Project Lead 15\% (9)   & 6 - 20 47\% (23) \\
                          & Middleware 52\% (26)      & Consultant 11\% (7)     & 21+ 37\% (18)    \\ \hline
\multirow{3}{*}{Critical} & Web apps 68\% (15)        & Programmer 41\% (9)     & 0-5 5\% (1)      \\
                          & Desktop apps 45\% (10)    & Project Lead 27\% (6)   & 6 - 20 45\% (10) \\
                          & Middleware 36\% (8)       & Consultant 9\% (2)      & 21+ 50\% (11)    \\ \hline
\multirow{3}{*}{ML}       & Deep learning 65\% (13)   & ML/DL Engineer 20\% (4) & 0-5 45\% (9)     \\
                          & Non-deep learning 5\% (1) & Researcher 20\% (4)     & 6-10 50\% (10)   \\
                          & Both 30\% (6)             & Data Scientist 15\% (3) & 11-15 5\% (1)    \\ \hline
\multirow{3}{*}{CPS}      & \multirow{3}{*}{-}        & Project Lead 17\% (1)   & 10-15 17\% (1)   \\
                          &                           & Researcher 17\% (1)     & 16-20 67\% (4)   \\
                          &                           & Programmer 17\% (1)     & 21+ 17\% (1)     \\ \hline
Legal                     & -                         & -                       & 13              \\ \hline
\end{tabular}
}
\vspace{-0.3cm}
\end{table}

\subsection{Interviews Design and Response Analysis}

We conducted one-hour semi-structured interviews with eight participants of the surveys (see \Cref{tab:response_stats}), to gather deeper knowledge about their experience and responses. 


\rev{Upon agreeing to answer surveys, respondents indicated willingness to be contacted for follow-up interviews. We selected respondents from the 5 surveys whose responses warranted further investigation.
In particular, we sought interviews with respondents who (1) gave detailed replies highlighting interesting use cases, challenges, and potential solutions; (2) demonstrated experience in their field; and (3) diversified our interviewee pool in terms of their role (consumers, producers, \etc). 
We hoped to capture a variety of perspectives from respondents familiar with SBOMs and those that were not but had interesting thoughts on how SBOMs might affect them.
Potential interviewees were identified independently by researchers during the open-coding process, and 11 participants were contacted upon consensus. 
In total, 8 participants accepted and completed an interview (\Cref{tab:response_stats}).} 

The interviews were conducted in two parts. The first part asked follow-up and clarification questions which varied depending on the 
survey responses of each interviewee (\eg \textit{You highlight the importance of identifiers for each software element. Why are these identifiers so important?}). For interviewees in the SBOM community and adopters group, a second part of the interview featured five questions that were common across all interviews in that group.  They asked about general themes and trends observed in the survey which had a broad impact on stakeholders. Our replication package contains the protocol we followed for the interviews \cite{anonymous_repo}.

Interviews were conducted over Zoom and recorded with the participants' permission. The recordings were transcribed using the Whisper speech recognition tool \cite{radford2022robust}. The interviews included two authors, taking notes about the given responses. The same authors parsed and analyzed participant responses and notes individually, employing an open coding strategy like that used in the analysis of the survey responses and discussing the coding when needed. 

\rev{Interviewees were given a \$50 USD Amazon gift card.}


\section{Study Results}
\label{sec:results}



\rev{56\% (84/150)} of the study participants are familiar with SBOMs (see \Cref{tab:response_stats}). The 22 respondents from the "Critical" survey belong to 16 of the 102 (15.7\%) critical OSS.

\subsection{\ref{rq:usage}: SBOM Creation and Usage}

\subsubsection{SBOM awareness and formats} 

\rev{Of the 50 producers, consumers, and tool makers surveyed, 16 reported using SPDX, 8 CycloneDX, and 12 both. SWID~\cite{ntiaGlance} was used by only 5 respondents, often with other formats.}  Those that consume SBOM, do so frequently: \rev{35.5\% (11/31) of participants stated they use them daily and 29\% (9/31) weekly.}  
Of the 22 critical OSS survey participants, 9 were unfamiliar with SBOMs and 7 were aware of SBOMs, while not adopting them yet. 
One interviewee mentioned how the limited interest is also due to the limited tool support and the need for manually maintaining SBOMs \rev{(in line with Zahan \etal's findings \cite{zahan2023software}).}
\looseness=-1

It is possible that private organizations and closed-source projects use SBOMs---in any of the standard formats or their own---yet our study did not find any evidence of that. For example, it is known that CERN uses CycloneDX \cite{cern_1, cern_2} and popular standards have been mentioned by Eggers \etal for the nuclear industry \cite{eggers2022towards}.

Of 6 CPS respondents, 3 were familiar with HBOMs and 2 had used them, but with bespoke formats.

\rev{No ML practitioners surveyed were aware of BOM formats for AI systems or datasets, but one interviewed standard maker was on an SPDX team that worked on adding fields to SPDX 2.x for ML systems: fields for ``describing data, the data sources, the data owners who you receive the data from, like did you buy it? Did you get it from open source? What were the references for the data you used to train the model? If it’s available, also the pointer to the public information about the data”. At the time of writing this paper, we have also learned that 
CycloneDX has added a Machine Learning Bill of Materials (ML-BOM) to its specification \cite{cyclone_main}.}

\rev{Participants expressed that pressure to maintain SBOMs primarily targets industry and projects at the end of a supply chain, while projects near the beginning have little incentive to produce them.} Some projects, such as the Linux kernel, may have no real dependencies of their own and so do not require dependency management methods.  As one interviewee noted, "I don't see a rush to add SBOMs to the originating open source. I see a rush to add SBOMs to the middle folks..."

This results in downstream components creating SBOMs on behalf of their dependencies. Other than being a cumbersome task done for somebody else; as one interviewee said, "[the risk is] miss[ing] something because you got to go back and dig back through all these different dependencies."




\subsubsection{SBOM use cases, benefits, and data fields}


In line with existing SBOM documentation~\cite{ntia_benefits} and prior studies~\cite{xia2023empirical, zahan2023software,linuxSbomSurvey}, we found that security, dependency tracking \cite{rousseau2020software}, and licensing are the main use cases for SBOMs. 
Out of 61 SBOM practitioners, 55 mentioned as main use case dependency management, 22 licensing concerns, and 22 software security (\eg vulnerability) management.
Other responses include software versioning (14), provenance (10), documentation (6), and transparency (4).

While tracking vulnerabilities was a main use case for \rev{consumers~(80.7\%), producers (100\%), and tool makers (83.3\%),}
some respondents were concerned that SBOMs might provide a road map of vulnerabilities for attackers.  \rev{This misconception, also identified by Zahan \etal \cite{zahan2023software}, has been addressed by NTIA \cite{myths_and_facts} and our interviewees rejected the  notion of "security by obscurity."}

When 41 SBOM producers, tool makers, and standard makers were asked which data fields should be included in SBOMs, responses varied. The most common answers were general information about the software components:
version number (24 \rev{of 41}), license (22), component name (18), and a URL to the component (18). Notably, 13 respondents indicated that the SBOM should contain unique identifiers for the software component the SBOM is documenting and/or its dependencies~\cite{coswid, cpe, purl, heritage, gitoid_1,gitoid_spdx}.


\rev{Although we found little evidence to suggest AI and DataBOMs are being used in practice, respondents mentioned two potential use cases.  These BOMs could facilitate ML model reproducibility and help to identify / verify datasets across academic papers.  Specifically, AIBOMs can provide transparency into how a model was trained, providing information about its architecture, hyper-parameters, and any pre-trained base models used. By providing provenance and usage information, a linked DataBOM can also make developers aware if a model was trained using a poisoned, biased, or illegally sourced dataset.} 

\rev{When asked about the ideal relationship between AI and DataBOMs, 9 of 20 (45\%)  respondents stated they should be separate documents and 5 (25\%) that they should be complementary.  Only 2 (10\%) proposed that the documents should be combined.}

\rev{The surveyed and interviewed CPS practitioners mentioned that BOMs could serve as regulatory documents for critical embedded systems (consistent with the findings of \cite{carmody2021building}), and that they could increase the transparency and reproducibility of research results in academic communities. For these tasks, the BOMs must communicate information related to the physical hardware components (part numbers, manufacturer, \etc), firmware, and other software (including configurations) of the system.}

\subsubsection{SBOM generation process, tooling, and distribution}

There was little consistency in the tools used across participants, with there being a mix of in-house, commercial (\eg Anchore \cite{anchore}), and open source solutions (\eg ScanCode \cite{scancode}).  

\label{sec:dist}
Despite the NTIA recommendations \cite{sbom_Sharing}, there is currently no agreed-upon method for distributing SBOMs. 
Respondents have the expectation that the developers of third-party components they use should be the ones creating, maintaining, and distributing SBOMs along with their software. 
\rev{5 of 12 (41.67\%) critical OSS developers asked about SBOM deficiencies mentioned distribution as a challenge moving forward.}




Concerning support for DataBOMs and AIBOMs, two survey participants mentioned that Hugging Face dataset cards \cite{data_cards} could serve as DataBOMs.  
Three respondents mentioned the same service's model cards, providing similar information to  AIBOMs.  Other tools mentioned include DVC \cite{dvc} and ML-Flow \cite{ml_flow}.  \rev{These formats are ``quasi-AIBOMs'' since, to our knowledge, no formal AIBOM standards have been implemented and accepted in practice.}

When asked when SBOMs should be generated, producers said: during each build (28/34), when publishing a major release~(21/34), during deployment (19/34), and at the developer's discretion (7/34). 
\looseness=-1



\vspace{-0.28cm}
\subsection{\ref{rq:challenges}: SBOM Challenges}

We summarize and discuss the challenges of using and creating SBOMs, expressed by the participants.




\begin{enumerate}[label=(\textbf{C\arabic*}), wide, labelindent=5pt]

  \item \textbf{Complexity of SBOM  specifications.}
\label{sec:over_spec}
A common key concern among participants is the complexity of SBOM specifications, as stated in this comment: "[...] one core issue [...] is definitely a tension between use case coverage and the complexity of the spec." 
\looseness=-1


\rev{Adding support for new use cases lengthens and complicates SBOM specifications.}
A standard maker mentioned: 
"[They say,] `the spec's too complicated.  All I want to do is X. [...] You're missing something for X, so I want to add that in,' which makes it more complicated for the other 99 people [without that use case]." 

We noticed that the user's perception of the SBOM specification is in part determined by their use case.  
\rev{“If all you're interested in is licensing, [...] [you] don't want to have to learn [about other domains like security] just to be able to use the spec.”}
\rev{However, "even if [SBOM producers] don't have that use case in mind, [their] consumers [might]."}




Participants also mentioned the lack of adequate educational resources about the SBOM specifications to better communicate their content. One interviewee mentioned: "It's not just simplicity in the spec. It's not simplicity in the tooling, but how we message it and how we communicate it. Because if we send them to the [standard] spec website, they'll take a look at that and go, well, I'm not going through all that work".

  
  \item \textbf{Determining data fields to include in SBOMs.}
 \label{sec:fields} While some fields (software versions, licenses, or component names) are commonly agreed upon, others depend on the use case. For example, practitioners seeking to analyze their software for vulnerabilities may require BOMs to link to an external vulnerability database. 




Interesting is the case of BOMs for AI/ML.   \rev{AI/ML respondents expressed the need to include  provenance information about datasets and models in SBOMs, to enable model verification and reproducibility.}
Other than standard SBOM fields, 
the 20 respondents from this group pointed out fields such as descriptions of the training data (17) and validation/testing data (14), preprocessing steps taken on the data (13), dataset version (13), and used optimizers/loss functions (13). When asked about fields needed in DataBOMs, they highlighted data sources (18), data transformations (18), preprocessing steps (17), dataset size (16), known/potential biases (14), and data collection procedures (14).

\rev{Of the 6 surveyed CPS practitioners, 3 expressed a need for hardware part numbers, 2 for testing and quality assurance data, 1 for system deployment information, 1 for manufacturer information and location (\eg company and geographical location), and 1 for known limitations about parts (\eg if they are not suitable for certain tasks due to security risks).}

\rev{Adding additional fields to SBOM specifications makes the documents more useful, but as mentioned previously, also contributes to the complexity of the specification (C1).}

  
  \item \textbf{Incompatibility between SBOM standards.}
  \label{sec:incompat}
Responses show that competing standards confuse developers. When consuming SBOMs, 23.33\% of the SBOM practitioners stated that different standards pose a challenge, due to  interoperability issues between standards and inconsistency between standards and tooling.
\looseness=-1


Despite this, one practitioner said: "Competition is good [...] I definitely think that we have moved faster because of CycloneDX and SPDX having this kind of competition."  

There are also multiple ways of creating an SBOM for the same piece of software, often for backward compatibility reasons.  
One practitioner remarked: "You may have two SBOMs that technically represent the same software, but they're being produced by two different tools and they look radically different."  

Fortunately, respondents suggested there are plans to increase and maintain interoperability among different standards.  As one interviewee put it, "I think [the standards are] on two different paths now. [...] To say one's going to die over the other, or try to do the grand convergence and bring them together, you're just not going to, it's just going to take too long. [...] it makes much more sense to try to get the two groups to collaborate."

\rev{Addressing incompatibility between standards would likely require a community-led effort, creating clear mappings between them, and developing tools that support these mappings.}


  \item \textbf{Keeping SBOMs up to date.} \label{sec:update_sbom}
Once an SBOM has been created, it must be maintained along with the software it represents. Substantial changes to an SBOM over time are known as SBOM drift~\cite{sbom_drift_1}.  
\rev{Such changes can occur suddenly, such as a dramatic increase in the number of dependencies when an application is added to a container \cite{sbom_drift_2}, or when new vulnerabilities are discovered in dependencies asynchronously from changes in the software } 
--- one interviewee described SBOMs as "a static vulnerability snapshot of the state of a [piece of] software at a certain point of time." 

\rev{When asked about deficiencies in standards, 4.35\% of participants expressed issues concerning keeping SBOM updated (1), upkeep requirements (1), and the syncing of SBOM versions (1).}  Of 3 critical OSS developers that consume SBOM, 1 mentioned difficulties in keeping SBOMs up-to-date. This motivates a need for tools which can dynamically update SBOMs as changes occur \cite{dynamic_sbom}.






  \item \textbf{Insufficient SBOM tooling.} 
\label{sec:insuff_tooling}
Figure \ref{fig:tool_support} shows
stakeholders' views on whether current SBOM tools address the needs of their users. While we generally found a lack of consensus among participants, we observe that tool makers are slightly more negative. These results, combined with the participants' open-ended answers, suggest that current tool support is insufficient. One participant identified a lack of "automated ways to generate SBOM for embedded code like assembly, C, C++." 




\begin{figure}[t]
\centering
\includegraphics[width=0.92\linewidth]{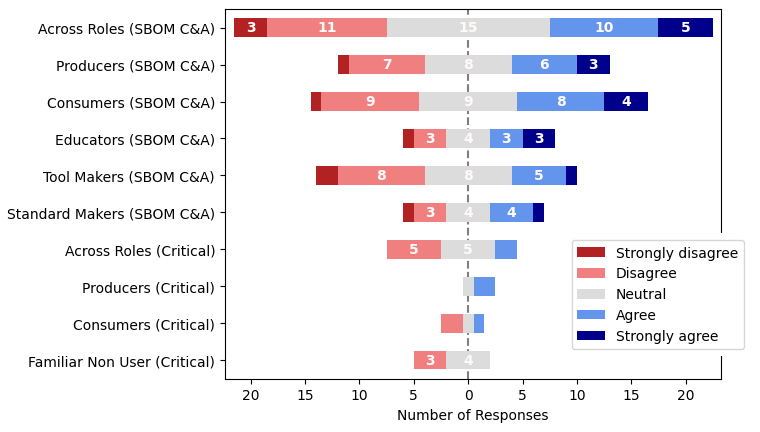}
\caption{Perceived sufficiency of SBOM tooling.}
\label{fig:tool_support}
\vspace{-.6cm}
\end{figure}

Across stakeholder groups, there was little familiarity with tools. 85\% of the ML respondents were unaware of any tool support for generating AIBOMs, and 90\% were unaware of tooling for DataBOMs.  Only one CPS practitioner was aware of existing tools. \rev{Part of the problem may be low demand. One practitioner had used "a few [SBOM] tools [but] they [didn't] work very well," noting that "it would be nice if they were fixed" but "nobody seems to care because maybe nobody's using them."}

Some projects with specific features may be unable to use current tooling, as no support exists for them yet. For example, \rev{one practitioner noted that current tooling could not "run fast on projects with tens of thousands of files...  They're not designed to work with very, very large projects."}  Two producers faced challenges involving projects that used multiple programming languages, 
suggesting an unmet need for tools to support multi-language projects.  Similarly, tools should be available for SBOMs to be created when only certain types of information are available, such as building SBOMs from binaries: "[T]here's source SBOMs. There's binary only SBOMs. There's SBOMs that have dependency information. There's SBOMs that have really just information about the package [...]." 



  \item \textbf{Inaccurate and incomplete SBOMs.}
\label{sec:poor_sbom}
 An SBOM is only as good as the information that it provides. If the information is inaccurate or incomplete, it becomes difficult for teams to make informed decisions concerning the dependencies, licensing, and security of their projects. 
 
 According to the results, currently available SBOMs are of varying quality and are often found wanting. 33\% of SBOM consumers from the SBOM C\&A survey  mentioned poor quality SBOMs as one of the challenges they had faced in using SBOMs. 25\% of the consumers from the critical OSS groups stated the same. Surprisingly, 12\% of the SBOM producers had the same complaint.

Consider that the minimum SBOM requirement would be to include all direct and \rev{transitive} dependency information, including the URLs of their sources. The legal practitioner we interviewed mentioned that, \rev{in his/her experience}, this condition is rarely met.


\rev{Participants also discussed "false positives" in BOMs.  For example, using a dependency that has a vulnerability does not necessarily mean the software will be impacted.  Determining if a project is actually impacted is a more difficult problem and requires more sophisticated tooling.}

The problem of inaccurate SBOMs also impacts tool developers. One respondent described how "it's been difficult to build tooling that accepts an SBOM when I'm not sure if all the fields that I'll need to depend on have been filled out."

  \item \textbf{Verifying SBOM accuracy and completeness.}
\label{sec:verify_sbom}
33\% of the critical OSS contributors mentioned how SBOM verifiability is a major challenge. This was also reported by 3 participants of the SBOM practitioner survey.
That being said, the enforcement of SBOM correctness should not be so strict that it impedes SBOM creation and adoption. For example, the legal practitioner we contacted cautioned that holding BOM creators liable for inaccuracies in the documents they produce is a disincentive to creating SBOMs at all. 

\rev{For security reasons, consumers will also need mechanisms to validate the integrity of an SBOM, to check that nobody has (maliciously) altered  it in transit. Well-known solutions, \eg those based on   hashing and checksums, can be applied to this context.}

  \item \textbf{Differences across ecosystems and communities.} \label{sec_differences}
\rev{Participants indicated that SBOM support varies across languages and package ecosystems.}
One interviewee mentioned: "a big part of the bottleneck is just retrieving all the information that needs to go into the SBOM and getting it from different sources [...] some language communities do a better job of capturing the metadata [to] include in the SBOM."  Some respondents even suggested that tools from the same standard (\eg CycloneDX) drastically vary in quality across languages. As another participant mentioned, this "creates an ecosystem challenge for getting that data in an SBOM in a reliable way, because there are some data sources that you can't really trust." 

We also observed challenges of creating SBOMs for languages with limited or no package managers. A survey respondent mentioned: "For C/C++ projects, dependencies are typically defined in autotools or cmake files, and Node, Ruby, Python, Golang, etc all have their own dependency management systems; typically recording exact versions is an output of the build process, although this doesn't come "out of the box" with C/C++ projects".

25\% of the critical OSS developers surveyed who were familiar with SBOMs listed a lack of language support as a deficiency in current SBOM specifications, while 8.7\% of SBOM practitioners agreed.  When asked about tool deficiencies, 41.67\% of critical OSS developers surveyed who were familiar with SBOMs expressed a need for more language-specific tooling.

  \item \textbf{SBOM completeness and data privacy trade-off.} \label{sec:privacy}
AI/ML participants indicated that AIBOMs and DataBOMs may entail a tradeoff between completeness and privacy on large datasets, given that these datasets may contain personally identifiable, private, sensitive, or proprietary information.  
CPS respondents also mentioned privacy concerns in BOMs, as CPS may actively collect and process private and sensitive data from the environment.  






\item \textbf{SBOMs for legacy packages and repositories.} One interviewee expressed the challenge of generating SBOMs for legacy software, which may be deployed and used by certain user groups.
\label{c:legacy}
Even if SBOMs become well-adopted and automatically generated during software builds, the question of what to do about legacy software remains.  Software that is still regularly maintained could feasibly have an SBOM created, but it is more challenging for older systems where the original source code is missing or for systems written in languages that are now substantially less common (\eg COBOL). These languages are less likely to be supported by open-source SBOM tooling.  This is particularly problematic for entities like the US government \cite{legacy_in_gov} or the banking industry \cite{limaj2020facing}. \rev{Community-driven effort may be needed to generate, store, and share SBOMs in such situations.}

An important question is, whether, for existing systems, only the newest releases require an SBOM, or if older releases that are still used by dependents also require SBOMs.  The respondent said: "if ecosystems did start to publish SBOMs, [...] it would be great to see [centralized repository maintainers] go back in time, generate SBOMs for older packages"

\item \textbf{Inability to locate dependencies for SBOMs.} 
\label{c:inabillity_to_locate}
There may be cases where during the production or consumption of an SBOM, a certain dependency cannot be located.  This could happen if a dependency was removed from a package manager (perhaps it was malicious or no longer maintained) or from the associated repository.  One practitioner mentioned: "They [dependencies] may have been yanked and removed from the upstream package registries, meaning that the mere fact of detecting that they exist could be a challenge" and "In some cases, [finding your dependency is] a lost cause in the sense that your source may be dead, the repository has disappeared and you're left to have to sift through random snapshots of archive.org calls made on the website. That's rare, but that happens."  Previous work shows that malicious packages exist \cite{awssecrets, npmpackageshacker, pypimaliciousimage, pypiw4sp, bagmar2021know, chen2016following, top_5_challenges, zahan2021weak, liang2021malicious, sejfia2022practical} and are commonly removed from package managers once detected \cite{tenpackagesremoved}.  Since CVEs are for vulnerabilities \cite{whatiscve}, entries for malware are not typically created, potentially leaving developers with a dead dependency reference and little way to discover the security threat the dependency poses.


\rev{A centralized database indexed on global IDs and containing provenance information for software repositories / distributions could allow developers to access critical information for projects that are no longer hosted or available.  This would essentially be a third-party SBOM archive.}

\item \textbf{Unclear SBOM direction and low adoption.}
\label{sec:unclear_direction}
\rev{A recent US executive order \cite{execOrder} requires companies selling software to the US government to provide corresponding SBOMs. While this has created incentives  to create and maintain SBOMs, our results indicate  how the adoption and knowledge of SBOM are still fairly limited. Moreover, while incentives for library users are clear, those for library creators are not. 
Therefore, given the effort and knowledge needed for creating SBOMs, most developers forgo this effort.}

The Information Technology Industry Council \cite{iti} wrote an open letter in response to recent pushes from the US federal government to mandate SBOMs~\cite{object_to_sbom}.  They assert that SBOMs are not yet suitable contract requirements: "The presence of multiple, at times inconsistent or even contradictory, efforts suggests a lacking maturity of SBOMs."  
They also raise concerns about cloud services, legacy software \ref{c:legacy}, and the protection of confidential or proprietary information \ref{sec:privacy}, all issues mentioned by respondents during our study. Though, many of these concerns have also been addressed by the NTIA \cite{myths_and_facts}.

This suggests a fear that the work required to create and maintain SBOMs will outweigh their benefits. As one of our practitioners said, "I hope that the hype around SBOM will lead to something that's productive [...] and will not just be something which is a compliance requirement that's going to be met in a minimal way.”  This fear was shared by  practitioners across domains.  Across our surveys, three respondents expressed worry that SBOMs would not be useful and another three feared that they would be time-consuming. 

Lastly, as we were reminded by numerous respondents, SBOMs are still not-mature-yet technology that will take time to mature.
Currently, there is still a need to motivate and implement support for consumer use cases.  
In an interview, one respondent stated, “You know, if you are a large organization and, say, you take a magic wand, and tomorrow all your software vendors start to provide accurate SBOMs, what are you going to do with this?”  

\end{enumerate}

\vspace{-.3cm}
\subsection{\ref{rq:solutions}: Solutions to SBOM Challenges}

\begin{table*}
\centering
\caption{How each SBOM solution addresses the SBOM challenges and the roles impacted by the challenges/solutions}
\setlength{\tabcolsep}{3pt}
\label{tab:solutions}
\resizebox{\textwidth}{!}{%
\begin{tabular}{lll}
\hline
\rowcolor[HTML]{EFEFEF}
\multicolumn{3}{|c|}{
 \textbf{(S1) Multi-dimensional SBOM specifications}} \\ \hline
\multicolumn{3}{|l|}{
\begin{tabular}[c]{@{}l@{}}
- Structure SBOM specifications considering three dimensions, i.e., use cases, types of software, and amount of information needed (\aka information level) \\ - Create more structured, easy-to-navigate, and easy-to-search specifications\\ - Improve educational material about SBOM specifications\end{tabular}} \\ \hline
\multicolumn{1}{|c|}{\textbf{Challenge}} & \multicolumn{1}{c|}{\textbf{How the Solution Addresses the Challenge}} & \multicolumn{1}{c|}{\textbf{Roles}} \\ \hline
\multicolumn{1}{|l|}{(C1) Complexity of SBOM specifications} & \multicolumn{1}{l|}{\begin{tabular}[c]{@{}l@{}}Shorter and easy-to-browse SBOMs specs. without unneeded information (per use case, system, \etc) 
\end{tabular}} & \multicolumn{1}{l|}{
\begin{tabular}[c]{@{}l@{}}
P, C, TM, E, SM
\end{tabular}
} \\ \hline
\multicolumn{1}{|l|}{(C2) Determining data fields to include in SBOMs} & \multicolumn{1}{l|}{\begin{tabular}[c]{@{}l@{}}Optional SBOM fields added only when required.\\
Information levels determine optional, recommended, and mandatory SBOM fields.\end{tabular}} & \multicolumn{1}{l|}{
\begin{tabular}[c]{@{}l@{}}
P, C, TM, E, SM
\end{tabular}
} \\ \hline
\multicolumn{1}{|l|}{(C6) Inaccurate and incomplete SBOMs} & \multicolumn{1}{l|}{
\begin{tabular}[c]{@{}l@{}}
SBOMs not incomplete if irrelevant/hard-to-find info for a given use case is not required. \\
\end{tabular}} & \multicolumn{1}{l|}{P, C, TM} \\ \hline
\multicolumn{1}{|l|}{(C9) SBOM completeness and data privacy trade-off} & \multicolumn{1}{l|}{
\begin{tabular}[c]{@{}l@{}}
SBOMs can tailor the required fields for data privacy
according to defined information levels.
\end{tabular}} & \multicolumn{1}{l|}{
\begin{tabular}[c]{@{}l@{}}
P, C, TM, SM
\end{tabular}
} \\ \hline
\rowcolor[HTML]{EFEFEF} 
\multicolumn{3}{|c|}{
\textbf{(S2)  Enhanced SBOM tooling and build system support}} \\ \hline
\multicolumn{3}{|l|}{\begin{tabular}[c]{@{}l@{}}- Develop libraries and base infrastructure for SBOM production, consumption, and verification\\ - Develop SBOM tooling for binaries and programming languages with no package managers\\ - Integrate SBOM creation into build and continuous integration (CI) systems and AI/ML frameworks (TensorFlow, \etc)\end{tabular}} \\ \hline
\multicolumn{1}{|c|}{\textbf{Challenge}} & \multicolumn{1}{c|}{\textbf{How the Solution Addresses the Challenge}} & \multicolumn{1}{c|}{\textbf{Roles}} \\ \hline
\multicolumn{1}{|l|}{(C4) Keeping SBOMs up to date} & \multicolumn{1}{l|}{
\begin{tabular}[c]{@{}l@{}}
SBOM tools  compatible with build and CI/CD automation 
to create/update SBOMs at each build.
\end{tabular}} & \multicolumn{1}{l|}{P, C, TM} \\ \hline
\multicolumn{1}{|l|}{(C5) Insufficient SBOM tooling} & \multicolumn{1}{l|}{
\begin{tabular}[c]{@{}l@{}}
Improved SBOM tools with support for multiple programming languages and AI/ML frameworks.
\end{tabular}
} & \multicolumn{1}{l|}{P, C, TM} \\ \hline
\multicolumn{1}{|l|}{(C6) Inaccurate and incomplete SBOMs} & \multicolumn{1}{l|}{
\begin{tabular}[c]{@{}l@{}}
SBOM tools integrated with build automation and AI/ML frameworks
create SBOMs with \\ dependencies actually used in binaries,
releases, and AI/ML models.
\end{tabular}
} & \multicolumn{1}{l|}{P, C, TM} \\ \hline
\multicolumn{1}{|l|}{(C7) Verifying SBOM accuracy and completeness} & \multicolumn{1}{l|}{
\begin{tabular}[c]{@{}l@{}}
Tools to check that SBOMs created from
source code and binaries contain the same dependency info.
\end{tabular}
} & \multicolumn{1}{l|}{P, C, TM} \\ \hline
\multicolumn{1}{|l|}{(C8) Differences across ecosystems and communities} & \multicolumn{1}{l|}{
\begin{tabular}[c]{@{}l@{}}
Improved SBOM tools  would lead to increased SBOM
adoption across languages and ecosystems.
\end{tabular}
} & \multicolumn{1}{l|}{P, C, TM} \\ \hline
 \rowcolor[HTML]{EFEFEF} 
\multicolumn{3}{|c|}{
\textbf{(S3) Strategies for SBOM verification}} \\ \hline
\multicolumn{3}{|l|}{- Third-party (community-based) certification/verification of SBOMs} \\ \hline
\multicolumn{1}{|c|}{\textbf{Challenge}} & \multicolumn{1}{c|}{\textbf{How the Solution Addresses the Challenge}} & \multicolumn{1}{c|}{\textbf{Roles}} \\ \hline
\multicolumn{1}{|l|}{(C6) Inaccurate and incomplete SBOMs} & \multicolumn{1}{l|}{
\begin{tabular}[c]{@{}l@{}}
With verification mechanisms in place, certified SBOMs would be
more accurate and complete.
\end{tabular}
} & \multicolumn{1}{l|}{P, C, TM} \\ \hline
\multicolumn{1}{|l|}{(C7) Verifying SBOM accuracy and completeness} & \multicolumn{1}{l|}{
\begin{tabular}[c]{@{}l@{}}
Verifying and certifying  SBOM content leads to enhanced
accuracy and completeness.
\end{tabular}
} & \multicolumn{1}{l|}{P, C, TM} \\ \hline
 \rowcolor[HTML]{EFEFEF} 
\multicolumn{3}{|c|}{\textbf{
(S4) Increasing incentives for SBOM adoption}} \\ \hline
\multicolumn{3}{|l|}{\begin{tabular}[c]{@{}l@{}}- Create mandates to create and use SBOMs for different stakeholders\\ - Minimize the effort to create and maintain SBOMs (\eg by developing tools integrated with existing systems and processes)\\ - Increase motivation to develop (open-source) SBOM tooling (e.g., via integration and badging in code repositories such as GitHub)\\ - Promote SBOMs benefits/usage and improve educational materials (e.g., by promoting successful cases of SBOM usage and tooling)\end{tabular}} \\ \hline
\multicolumn{1}{|c|}{\textbf{Challenge}} & \multicolumn{1}{c|}{\textbf{How the Solution Addresses the Challenge}} & \multicolumn{1}{c|}{\textbf{Roles}} \\ \hline
\multicolumn{1}{|l|}{(C5) Insufficient SBOM tooling} & \multicolumn{1}{l|}{
\begin{tabular}[c]{@{}l@{}}
Increased incentives for SBOM adoption would drive
further development of SBOM tooling.
\end{tabular}
} & \multicolumn{1}{l|}{
\begin{tabular}[c]{@{}l@{}}
P, C, TM, E, SM
\end{tabular}
} \\ \hline
\multicolumn{1}{|l|}{(C12) Unclear SBOM direction} & \multicolumn{1}{l|}{\begin{tabular}[c]{@{}l@{}}SBOM mandates and promotion/education materials
clarify SBOM benefits and usage costs. \\ 
SBOM incentives would better involve open-source communities in
SBOM creation/usage/promotion.\\
\end{tabular}} & \multicolumn{1}{l|}{
\begin{tabular}[c]{@{}l@{}}
P, C, TM,
E, SM
\end{tabular}
} \\ \hline
\end{tabular}%
}
\end{table*}

In this section, we discuss solutions for the identified challenges.
The proposed solutions do not address \ref{sec:incompat}, \ref{c:legacy}, \&~\ref{c:inabillity_to_locate}: \rev{these  require additional research to mitigate effectively, but insights and potential directions are discussed in their challenge descriptions.} \finalrev{\Cref{tab:solutions} provides a summary of the proposed solutions.}
\looseness=-1


\begin{enumerate}[label=(\textbf{S\arabic*}), wide, labelindent=5pt]

    \item \textbf{Multi-dimensional SBOM specifications.}
We identify three dimensions that contribute to the complexity SBOM specification: (1) the intended use case of an SBOM, (2) the type of software the SBOM is generated for, and (3) the amount of information documented in an SBOM.  Providing clear guidance for these dimensions is needed to inform consumers/producers which fields an SBOM should contain \ref{sec:fields}. The ultimate goal is to reduce the cognitive load placed on users of the specification \ref{sec:over_spec}. 


\underline{SBOM use cases.} 
\rev{As discussed, dozens of potential use cases exist for SBOMs~\cite{spdx_use_cases}, but including fields tailored for each of these results in cluttered specifications (see \ref{sec:over_spec}).} In interviews, we learned that the SPDX team is working on \textit{profiles} \cite{spdx_why_security} which define the fields required in an SBOM document meant for a specific use case. This will allow producers to create SBOMs tailored to their use case without worrying about irrelevant fields to them.  One practitioner mentioned that "being able to call [use case] out in these profiles will make [what to expect in the quality] a lot clearer. And I think that might help with [poor quality SBOMs] \ref{sec:poor_sbom}. Not so much making the quality of the SBOMs better, but at least making it obvious what the quality is."  Another said: "Let's say I want to just graph the relationships, right? There's a lot of data that's included in the SBOM that I wouldn't necessarily need. And if some of that data is expensive to calculate, then the tool that gives me the SBOM would run a lot faster if all I was ever looking for was a way to kind of graph the relationships." 

\underline{Types of software.} Different types of software require different information to adequately describe them.  ML-related software requires fields that firmware or cloud services likely will not.  Even though all three fall under the umbrella of software, it may be prudent to separate them into distinct SBOM types (AIBOM, FBOM, SaaSBOM, \etc), so that it is easier for end users to know the type of system the SBOM describes. This model of different SBOM types has already been adopted by CycloneDX~\cite{cyclone_capabilities}.



\underline{Amount of information in SBOMs.} Within the same use case and software type, users may desire different amounts of data in an SBOM.  One practitioner noted:
"it would be interesting to have different levels [...] where this has `level 1' data. [...] This tool generates `level 2' data, this tool generates `level 3' data...".
These data levels reflect the amount of information a user can expect to find in the SBOM.  \rev{Lower data levels could potentially be used for privacy-sensitive applications \ref{sec:privacy}.}
Data levels could also create some standardization in tooling: 
"I think it would help people who are writing tools [...] to be able to then differentiate between the level of data that they can expect to see within the SBOM."



Adding this flexibility to standards does not necessarily make them more complex or difficult to use. One practitioner indicated that "even though the minimum requirements that have been provided [...] seem to be or could be construed as daunting, the essence of what needs to be provided in SBOM can be surprisingly simple." Educational resources and documentation will have to be well-crafted to explain this approach.  


    
    \item \textbf{Enhanced SBOM tooling and build system support.}
Across all surveys, three respondents suggested better libraries as a tooling solution.  One said, "Increased investment in open source libraries that can be incorporated in end user commercial and open source tools [can address current deficiencies in tooling]."  \rev{Well-maintained, easy-to-use libraries would serve as the foundation and motivation to develop SBOM tools \ref{sec:insuff_tooling} providing functionality for creating, maintaining \ref{sec:update_sbom}, parsing, and managing SBOMs, enhancing the user experience and, potentially, SBOM adoption. }

Our findings indicate the need for language-specific SBOM production tools. A language-agnostic tool is unlikely to adequately support all scenarios. As such, there is work to be done creating SBOM generation tools for different ecosystems, including resolving disparities in the quality of available tools. Creating better tools will be a community effort: "part of it is just [...] being willing to get in and help out with the quality of those tools."  Language-specific tooling can be built on language-agnostic libraries \ref{sec_differences}. 

\rev{SBOMs will likely become more accurate and complete with better tool support \ref{sec:poor_sbom}. Respondents from the critical OSS survey pointed out that quasi-SBOM files are typically accurate  and are generated/checked automatically by tools: mature SBOM tools would likely be able to perform similarly.}

\rev{Moreover, in the current landscape of varying SBOM quality, consumption tools may also be responsible for checking the accuracy of the SBOMs consumed \ref{sec:verify_sbom}.  A respondent noted that consumption tools "have a perhaps harder job to make sure that the data that's being generated is accurate."} 

\rev{Furthermore, existing build systems (\eg Maven or Gradle) should be made SBOM-aware: capable of reading and generating SBOMs: "[O]ne way [for SBOMs to be easier to use] would be for build tools to start generating them without asking." We have observed from our surveys that developers tend to prefer processes or tools that are commonly used or predetermined: "when the recommended way of doing something is the default, then it gets done more often." SBOM generation functionality in build tools would more easily facilitate the update of SBOMs \ref{sec:update_sbom}.}


We have seen that developers rely on package management systems to obtain a list of their project's dependencies.  Many of these systems also provide quasi-SBOM files.  If SBOM generation and acquisition could be handled at the package manager level, we would likely see a large uptick in adoption \ref{sec:unclear_direction}.  SBOMs could be stored along with other package information and queried through APIs.  
Indeed, interviewed  practitioners suggested that SBOMs should be kept as close to the source as possible.  As an SBOM moves further from the source, it is less likely to be up-to-date \ref{sec:update_sbom}.

GitHub recently unveiled new functionality capable of generating SPDX documents for a cloud repository \cite{github_tool}. Through integration with GitHub's Dependency Graph tool \cite{dependency_graph}, this capability supports SBOM generation for a number of popular languages and is easily accessible to developers, marking a strong start for SBOM integration.
\looseness=-1



It was also suggested that ML libraries could generate AIBOMs or play an integral part in easily accessing \rev{required information: 
"eventually} there'll be [...] something built into TensorFlow or PyTorch
[...] that outputs a document 
\rev{[...]} that tells you the
key elements [like] the hyper-parameters."



    \item \textbf{Strategies for SBOM verification.}
\rev{One initially apparent method to approach incomplete or incorrect SBOMs would be to hold parties accountable for the SBOMs they generate \ref{sec:poor_sbom}, but this could lead to unintended consequences.  A legal practitioner said, "[a] requirement for them to certify that it is complete or correct is only going to result in fear of creating SBOMs. `Perfect' should not be the enemy of `good.'" Beyond this,} SPDX SBOMs are licensed under Creative Commons 0 (CC0) \cite{spdx_data_license, creative_commons_0}, meaning no warranty is included and the producer assumes no liability.  The open-source licensing of tools protects their creators from litigation since many licenses also do not provide a warranty \cite{mit}.  According to the legal practitioner we interviewed, issues of liability would likely only arise if proprietary software or service provided a warranty. He/she "could see there being contractually accepted liability as part of [one party agreeing to provide an accurate SBOM]." 

Two other solutions emerged from our surveys \ref{sec:verify_sbom}.  A third-party certification or review board could approve SBOMs and endorse them. \rev{However, as one respondent put it, "central authorities have never seemed to work too well in our industry [...]". An alternative, decentralized approach could involve the assessment of SBOMs by their consumers and other stakeholders, with issues reported to the SBOM producer or posted in a shared database.}


    \item \textbf{Increasing incentives for SBOM adoption.} 
There is a need to either minimize the effort needed to
create and maintain SBOMs (such as improving current development toolkits to generate them) or by gaining other benefits, such as having tools that consume SBOM and help with developer tasks. Similarly, it is
necessary to motivate the creators of the development toolkits to support SBOM creation \ref{sec:insuff_tooling}. Github's new SBOM tools are a
step in the right direction. Also, issuing badges might be a simple incentive that might promote the adoption of SBOMs (as it has been in other domains \cite{DBLP:conf/icse/TrockmanZKV18}) \ref{sec:unclear_direction}.

Similar to Executive Order 14028, other stakeholders could require their participants to provide SBOMs. For example, the scientific publication of tools and models could require that artifacts be accompanied by SBOMs \ref{sec:unclear_direction}. These SBOMs would increase the transparency of the work and ideally increase reproducibility.

At the same time, better marketing and educational materials that emphasize the importance of SBOMs are needed, \rev{both for software developers and consumers.} 
As one user put it, "It’s not just simplicity in the spec [nor] simplicity in the tooling, but how we message it and how we communicate it."

Ultimately, creating and using SBOMs should be done because it helps to create and maintain better, more secure, and reliable software, and that ultimately benefits society.

\end{enumerate}

\vspace{-0.3cm}

\section{Threats to Validity}

\begin{enumerate}[label={}, wide, labelindent=5pt]

\item \textbf{External Validity.} 
The conclusions of our study apply to the population that participated in the survey and interviews. \rev{By design, we cannot overly generalize our results \cite{baltes2016worse}, yet our observations pertaining to open-source developers may extend to other  open-source projects.  Generalizability for the industry is more difficult, but industries within the same country will abide by the same legislation and regulations, likely resulting in similar use cases and challenges.  Ultimately, our goal was not to claim generalizability, but to gain a clearer understanding of the current landscape of SBOM usage, the challenges therein, and how to overcome them. While the number of respondents for the ML, CPS, critical, and legal surveys is rather small, they provided insights from the perspectives of \rev{practitioners} (belonging to different areas) who may or may not use SBOMs firsthand, which are still valuable to understand the current landscape and future directions of SBOMs.} 

\item \textbf{Internal Validity.} 
To mitigate researcher bias in open-ended response coding, we followed an iterative, hybrid coding process that included discussion for all disagreements to reach a consensus such that the codes applied to a given response most accurately reflected its content.  
To ensure that we surveyed practitioners with different backgrounds, we employed a diverse set of strategies to find participants, including the search for relevant repositories on GitHub, posting to relevant mailing lists, and contacting practitioners through our professional network. \rev{However, the low response rate and self-selection bias may have influenced the results by attracting participants interested in the survey topic.} 
We formulated our survey/interview questions to follow best practices and survey/interview design guidelines. We ensured questions were clear and concise, avoiding language that would bias respondents towards a certain answer, and providing clarification and defining terms we used when necessary.
Additionally, we mitigated potential confirmation bias in our qualitative analysis by performing independent coding, discussing disagreements, and reaching a consensus 
backed with facts from the data. 
While we attempted to remove AI-generated responses from results, they remain a well-accepted risk in this kind of study. 

\end{enumerate}

\vspace{-.25cm}

\section{Conclusion}
\label{sec:conclusion}
This paper reports and discusses the findings from a study---conducted through surveys and interviews with \rev{software practitioners}
---on the use of bills of materials for software systems. Other than targeting a general population of SBOM adopters, we also targeted specialized populations of
developers of critical OSS, as well as 
AI/ML, CPS, and legal practitioners.

The study results indicate that, while the adoption of SBOMs is still low, practitioners utilize them in a variety of use cases at various stages of software development and maintenance, including software licensing, dependency management, and security assessment. 
While SBOMs have the potential to aid in both research and industry, tool support and SBOM standards are nearly nonexistent in specific areas such as AI/ML and CPS.

The wide variety of use cases for SBOMs, and the complexity and heterogeneity of software systems, have  led to numerous challenges, such as the complexity of standard specifications, inadequate tooling, or data privacy vs. completeness tradeoffs. 
To address such challenges, our study has identified a number of solutions and opened the road for future research and development in this area.

\vspace{-0.3cm}
\section*{Data Availability}

We provide an anonymized replication package containing survey and interview protocols, aggregated results, a code catalog for survey and interview responses with definitions, code to process results, and other data required for verifiability~\cite{anonymous_repo}.

\vspace{-.3cm}

\section*{Acknowledgements}
\label{sec:acknowledgements}
\finalrev{We thank the study participants for their time and valuable contributions.  This research was partially funded by NSF CCF-2217733.  A complete, detailed list of image attributions can be found at  \cite{anonymous_repo}.}

\balance

\bibliographystyle{ACM-Reference-Format}
\bibliography{references}

\end{document}